\documentclass[runningheads]{llncs}

\usepackage{multicol}
\usepackage{xspace}
\usepackage{enumitem}
\usepackage{tikz}
\usetikzlibrary{tikzmark}
\usetikzlibrary{positioning}
\usepackage{pdflscape}
\usepackage{hyperref}
\usepackage{doi}
\usepackage[T1]{fontenc}
\usepackage{todonotes}
\usepackage{csquotes}
\usepackage{mathpartir}
\usepackage{simplebnf}
\usepackage{amssymb}
\usepackage{amsfonts}
\usepackage{amsmath}
\usepackage[nameinlink]{cleveref}
\usepackage{stmaryrd}
\usepackage{color}
\usepackage{graphicx}

\usepackage{simplebnf}
\RenewDocumentCommand{\bnfexpr}{m}{\( #1 \)}
\RenewDocumentCommand{\bnfannot}{m}{\textit{(#1)}}
\RenewDocumentCommand{\SimpleBNFDefOr}{}{\;\ensuremath{|}\;}

\NewCommandCopy{\oldbnfgrammar}{\bnfgrammar}
\NewCommandCopy{\endoldbnfgrammar}{\endbnfgrammar}

\RenewDocumentEnvironment{bnfgrammar}{O{lrclr} O{\|\|} O{\|}}
{\begin{oldbnfgrammar}[#1][#2][#3]}
{\end{oldbnfgrammar}}

\SetBNFConfig{
        or-sym = {$ \ | \ $}
}

\SetBNFLayout{
        colspec = {llcl},
        column{1} = {font = \sffamily},
        column{2} = {mode = dmath},
        column{4} = {mode = dmath},
}

\usepackage{semantic}
\mathlig{!->}{\mapsto}
\mathlig{|->}{\mapsto}
\mathlig{<-|}{\mapsfrom}
\mathlig{|->*}{\mapsto^{*}}
\mathlig{|[}{\llbracket}
\mathlig{|]}{\rrbracket}
\mathlig{|>}{\vartriangleright}
\mathlig{|>*}{\vartriangleright^{*}}
\mathlig{<|}{\vartriangleleft}
\mathlig{=ciu}{\overset{ciu}{\approx}}
\mathlig{=ctx}{\overset{ctx}{\approx}}
\mathlig{=?=}{\overset{?}{=}}
\mathlig{=[}{\sqsubseteq}
\mathlig{=]}{\sqsupseteq}

\newcommand{\ie}{\textit{i.e.,}\xspace}

\newcommand{\defeq}{\overset{\textit{def}}{=}}

\newcommand{\judgshape}[2][]{\begin{flushleft}\fbox{\(#2\)}~#1\end{flushleft}}

\newcommand{\mmathbf}[1]{\ensuremath{\mathbf{#1}}}
\newcommand{\subst}[3]{{#1}[{#2}/{#3}]}

\newcommand{\tyjudg}[3]{#1 \vdash #2 : #3}

\newcommand{\conf}[2]{\langle #1 \;\ensuremath{|}\; #2 \rangle}
\newcommand{\step}{\rightarrow}
\newcommand{\stepstar}{\step^{*}}

\newcommand{\hwt}[2]{#1 \; : \; #2}
\newcommand{\lw}[2]{\lfloor #1 \rfloor_{#2}}
\newcommand{\lf}[1]{\lceil #1 \rceil}

\newcommand{\lang}{$\lambda_{\text{PR}}$\xspace}

\newcommand{\lam}[3]{\<lam> #1 : #2 . #3}
\newcommand{\app}[2]{#1 \; #2}
\newcommand{\funty}[3]{#1 \xrightarrow[]{#3} #2}
\newcommand{\new}[1]{\<new> #1}
\newcommand{\deref}[1]{\<deref> #1}
\newcommand{\refty}[1]{\<Ref> #1}
\newcommand{\unite}{\<unit>}

\reservestyle{\asmc}{\mmathbf}
\asmc{deref[\boldsymbol{!}\;],
  new[new\;],
  nsnew[new],
  nsderef[\boldsymbol{!}],
  let[let\;],
  in[\;in\;],
  lam[\boldsymbol{\lambda}\;],
  Ref[Ref\;],Type[Type],max[max\text{-}level],
  Nat,
  Bool,
  Unit,unit[\langle \rangle]
}

\begin{document}

\title{Type Universes as Allocation Effects}

\author{Paulette Koronkevich
  \and
  William J. Bowman
}

\institute{University of British Columbia\\
\email{pletrec@cs.ubc.ca}\\
\email{wjb@williamjbowman.com}}

\maketitle              

\begin{abstract}
In this paper, we explore a connection between type universes and memory
allocation.
Type universe hierarchies are used in dependent type theories to ensure
consistency, by forbidding a type from quantifying over all types.
Instead, the types of types (universes) form a hierarchy, and a type can only
quantify over types in other universes (with some exceptions), restricting
cyclic reasoning in proofs.
We present a perspective where universes also describe \emph{where} values are
allocated in the heap, and the choice of universe algebra imposes a structure
on the heap overall.
The resulting type system provides a simple declarative system for reasoning
about and restricting memory allocation, without reasoning about reads or writes.
We present a theoretical framework for equipping a type system with
higher-order references restricted by a universe hierarchy, and conjecture that
many existing universe algebras give rise to interesting systems for
reasoning about allocation.
We present 3 instantiations of this approach to enable reasoning about
allocation in the simply typed $\lambda$-calculus: (1) the
standard ramified universe hierarchy, which we prove guarantees termination of
the language extended with higher-order references by restricting cycles in the
heap; (2) an extension with an \emph{impredicative} base universe, which we
conjecture enables full-ground references (with terminating computation but
cyclic ground data structures); (3) an extension with \emph{universe
polymorphism}, which divides the heap into fine-grained regions.
(This is a fresh perspectives submission.)

\end{abstract}

\section{Introduction}
Many type systems have been designed to reason about memory.
But what does it mean to reason about memory?

A large amount of work is dedicated to \emph{safety}, ensuring updates,
aliasing, \emph{etc.}, do not cause bugs when reading and writing from memory.
For example, L3~\cite{ahmed2007} has a type system that enforces safety with strong updates,
\ie updates that may change the type of already allocated locations on the heap.
L3's type system focuses on the safety of reads in the presence of these strong
updates, and thus focuses solely on the problem of aliasing.
Rust~\cite{matsakis2014} implements a similar system, enabling reasoning about
aliasing in the presence of concurrency.

Some type systems enable reasoning about \emph{deallocation}.
Region type-and-effect systems~\cite{tofte1997} were originally designed
for static fine-grained memory management, and Rust follows in this tradition.

Much of this work reasons about memory \emph{indirectly}, from how references to memory are used.
In Rust, allocation and deallocation is inferred from usage---when a value is
created, or copied, it is allocated, and it is deallocated when the value is no longer owned.
Regions are also inferred through individual reference reads and writes.
Any heap structure, such as the stack of regions structure~\cite{tofte1997} or
stratified regions~\cite{boudol2010,amadio2009}, is also inferred through reads
and writes.
L3 turns out to be terminating, which is enforced indirectly because of the
linearity usage requirement on references, but not because of any direct
restriction on structure of the heap.

We present a theoretical framework for designing type systems to enable
\emph{directly} reasoning about \emph{allocation}, by giving a declarative
description of allocation.
By declarative, we mean that each type \emph{declares} where values of this
type are allocated on the heap, and parameters of the type system
\emph{declare} what heap shapes and dependencies are allowed, in contrast to
these other type systems where allocation is \emph{inferred} from usage of heap
references.
For the time being, we do not consider reasoning about aliasing, although we
conjecture that existing designs that track aliasing could be integrated into
our type systems.

Our framework is based on a \emph{type universe hierarchy}, where the
underlying hierarchy can be changed to result in different heap structures.
Type universes are used to eliminate inconsistencies in dependent type
theories, which can arise when the type of types (the \emph{universe} \<Type>),
has type \<Type>.
This, and similar patterns, can lead to inconsistencies by admitting cyclic
proofs.
A standard solution to this is to stratify universes, each one a member of the next.
Instead of considering the \<Type> to be in universe \<Type>, dependent type
theories use a universe hierarchy to assign a level to each universe, so
$\<Type>_i$ is in universe $\<Type>_{i+1}$, ensuring consistency.
In our framework, each type has a universe, but the universe describes where
values are allocated, and the universe hierarchy enforces a heap structure for
all programs.
By changing the universe hierarchy, or level algebra, we impose different heap structures.

We first present an instantiation of our framework with a standard predicative
universe hierarchy, which enforces a stratified, acyclic heap.
Because the heap is acyclic, we prove this language with higher-order references is
terminating.
The language is simplified compared to similar work using stratified regions via a
type-and-effect system~\cite{boudol2010,amadio2009}.
Our language does not require tracking of reads or writes of individual
references, nor region inference.
We also present the proof of termination, which relies on a semantic notion of
garbage collection based on universe level.
We conjecture that the semantics might be lead to a declarative syntax for
deallocation.

We then abstract this type system to present the main parameters that change
the underlying universe algebra to create different heap structures.
We give two instantiations of this framework with existing type universe
hierarchies, resulting in type systems that enforce different heap structures.
In particular, we design a language with a heap with one level where cycles can
occur, based on universe hierarchies with a single \emph{impredicative} base
universe.
We conjecture the language is still terminating and yields
full-ground references~\cite{murawski2012}.
We also explore an extension with \emph{universe polymorphism}, which allows
a user to name regions of the heap and results in a system with more
fine-grained regions and conjecture could be used for static memory
management.

In short, we present a fresh perspective on viewing type universes as a
system for \emph{simple}, \emph{direct}, \emph{declarative} reasoning about
allocation---when and where allocation occurs, and which allocations a
computation depends on.

\newcommand{\FigSyntax}[1][t]{
\begin{figure}[#1]
      \begin{bnf}
        \tau : Type ::=
        | \<Unit>
        // \<Nat>
        // \funty{\tau}{\tau}{k}
        // \refty{\tau}
        ;;
        e : Expr ::=
        | x
        // n
        // \unite
        // \lam{x}{\tau}{e}
        // \app{e}{e}
        // \new{e}
        // \deref{e}
        // e \coloneqq e
      \end{bnf}
      \caption{\lang syntax}
      \label{fig:syntax}
\end{figure}
}

\newcommand{\FigTyping}[1][t]{
  \begin{figure}[#1]
    \judgshape{\tyjudg{\Gamma}{e}{\tau}}
    \begin{mathpar}
      \inferrule
        { }
        {\tyjudg{\Gamma}{n}{\<Nat>}}

      \inferrule
        { }
        {\tyjudg{\Gamma}{\unite}{\<Unit>}}

      \inferrule
        {x : \tau \in \Gamma}
        {\tyjudg{\Gamma}{x}{\tau}}

      \inferrule
        {\tyjudg{\Gamma,x:\tau_1}{e}{\tau_2} \\
        \colorbox{gray}{$k \geq \<max>(\Gamma,\tau_1,\tau_2)$}}
        {\tyjudg{\Gamma}{\lam{x}{\tau_1}{e}}{\funty{\tau_1}{\tau_2}{k}}}

      \inferrule
        {\tyjudg{\Gamma}{e_1}{\funty{\tau_1}{\tau_2}{k}} \\
        \tyjudg{\Gamma}{e_2}{\tau_1}}
        {\tyjudg{\Gamma}{\app{e_1}{e_2}}{\tau_2}}

      \inferrule
        {\tyjudg{\Gamma}{e}{\tau}}
        {\tyjudg{\Gamma}{\new{e}}{\refty{\tau}}}

      \inferrule
        {\tyjudg{\Gamma}{e}{\refty{\tau}}}
        {\tyjudg{\Gamma}{\deref{e}}{\tau}}

      \inferrule
        {\tyjudg{\Gamma}{e_1}{\refty{\tau}} \\
        \tyjudg{\Gamma}{e_2}{\tau}}
        {\tyjudg{\Gamma}{e_1 \coloneq e_2}{\<Unit>}}
    \end{mathpar}
    \caption{\lang typing.}
    \label{fig:typing}
  \end{figure}
}

\newcommand{\FigInfer}[1][t]{
  \begin{figure}[#1]
    \judgshape{\tau :: \<Type>_i}
    \begin{mathpar}
      \inferrule
        { }
        {\<Nat> :: \<Type>_0}

      \inferrule
        { }
        {\<Unit> :: \<Type>_0}

      \inferrule
        {\tau :: \<Type>_i}
        {\refty{\tau} :: \<Type>_{i+1}}

      \inferrule
        {\tau_1 :: \<Type>_i \\
        \tau_2 :: \<Type>_j \\
        k \geq \<max>(\tau_1,\tau_2)}
        {\funty{\tau_1}{\tau_2}{k}  :: \<Type>_k}
    \end{mathpar}
    \caption{\lang kinding.}
    \label{fig:kind}
  \end{figure}
}

\newcommand{\FigAuxDefs}[1][t]{
  \begin{figure}[#1]
    \begin{displaymath}
      \begin{array}{l}
        \textbf{World extension} \\
        W'_k =]_k W_k \iff \text{dom}(W_k) \subseteq \text{dom}(W'_k) \land \forall l \in W_k . W'_k(l) = W_k(l) \\
        \textbf{Lowering} \\
        \lw{W_k}{i} = \{ (l , \mathcal{V}|[ \tau |]_{j} ) \; | \; j < i \} \\
        \textbf{Lifting} \\
        \lf{W_k} = \lambda i . \Bigl\{\begin{array}{ll}
                                        \lambda l . W_k(l), & 0 < i \leq k\\
                                        \emptyset, & \mbox{otherwise}\end{array}
      \end{array}
    \end{displaymath}
    \caption{Additional definitions for worlds.}
    \label{fig:ad}
  \end{figure}
}

\newcommand{\FigVR}[1][t]{
  \begin{figure}[#1]
    \begin{displaymath}
      \begin{array}{lcll}
        \mathcal{V}|[\refty{\tau}|]_{i+1}(W_{i+1}) & \defeq & \{ l \; | \;
        W_{i+1}(l) = \mathcal{V}|[\tau|]_{i} \}
        \\
        \mathcal{V}|[\funty{\tau_i}{\sigma_j}{k}|]_{k}(W_{k}) & \defeq & \{
        \lam{x}{\tau_i}{e} \; | \; \forall W'_k =]_k W_k ,
        v \in \mathcal{V}|[\tau|]_i(\lw{W'_k}{i}) .
        \subst{e}{v}{x} \in \mathcal{E}|[\sigma|]_j(\lf{W'_k}) \}
      \end{array}
    \end{displaymath}
    \caption{Value relation for reference and function types.}
    \label{fig:vr}
  \end{figure}
}

\newcommand{\FigER}[1][t]{
  \begin{figure}[#1]
    \begin{displaymath}
      \begin{array}{rcll}
        \mathcal{E}|[\tau|]_i(\mathbf{W}) & \defeq & \{ e \; | \;
        \begin{array}[t]{l}
        \forall \mathbf{W'} =] \mathbf{W}, \hwt{h'}{\mathbf{W'}} .
        \conf{h'}{e} \stepstar \conf{h''}{v} \land \\
        \exists \mathbf{W''} =] \mathbf{W'} . \hwt{h''}{\mathbf{W''}} \land
        v \in \mathcal{V}|[\tau|](\lw{\mathbf{W''}}{i})\}
        \end{array}\ignorespacesafterend
      \end{array}
    \end{displaymath}
    \begin{displaymath}
      \begin{array}{rcll}
        \mathcal{G}|[ \cdot |](\mathbf{W}) & \defeq & \emptyset
        \\
        \mathcal{G}|[ \Gamma,x:\tau |](\mathbf{W}) & \defeq & \{ \gamma[x |-> v] \; | \;
        \tau :: \<Type>_i \land
        v \in \mathcal{V}|[ \tau |]_i(\lw{\mathbf{W}}{i}) \land
        \gamma \in \mathcal{G}|[ \Gamma |] (\mathbf{W}) \}
  \end{array}
    \end{displaymath}
    \caption{Expression and context relations.}
    \label{fig:er}
  \end{figure}
}

\section{Type Universes for Acyclic Heaps}
\label{sec:acyclic}
We now study how cycles in the heap can be created with
higher-order references through a pattern called
\emph{Landin's Knot}~\cite{landin1964}.
We show how to eliminate such cycles in a typed language equipped
with a universe type hierarchy.
In the type system we present, Landin’s Knot is ruled out in a purely
declarative way, without any reference usage tracking.

Preventing cycles in the heap is useful to achieve various language design and
implementation goals.
Reference counting, used to implement memory management, cannot handle (strong)
cycles in the heap~\cite{mcbeth1963}.
Certain cycles through the heap can lead to unrestricted recursion, which one
might instead desire to control through only explicit recursive constructs.
For example, dependent type theories require strong normalization for
decidability of type checking, shown for example by Jutting~\cite{jutting1993},
and termination might be desired to ensure for fairness in concurrent
settings~\cite{boudol2010}.

Unfortunately, the common wisdom is that higher-order references introduce
cycles in the heap that are difficult to prevent.
Landin's Knot, a pattern that uses a function capturing a mutable reference
and updates that reference to contain the function itself, is a typical example
of these kinds of cycles.
This pattern is illustrated through the following diverging program.
\begin{displaymath}
  \begin{array}[t]{ll}
    id = (\<lam> x . x) & : \<Nat> \to \<Nat> \\
    r = \<new> id & : \<Ref> (\<Nat> \to \<Nat>) \\
    f = (\<lam> x . (\<deref> r) \ x) & : \<Nat> \to \<Nat>  \\
    r := f; \\
    f \ 0
  \end{array}
\end{displaymath}

Three expressions $id$, $r$, and $f$ are defined in this program.
The expression $id$ is the identity function for natural numbers, and
$r$ is a reference initialized with $id$.
The second function $f$ captures the reference $r$, then expects a natural
number $x$ and and applies the function stored in $r$ to $x$.
The program diverges because it updates the reference $r$ to store the function
$f$, enabling unrestricted recursion in $f$.
The function $f$ calls the function stored in $r$, which is now $f$ itself, and
the program diverges.

We visualize the heap and dependencies on mutable references to better illustrate
the cause of this cycle.
In the following diagram, the reference $r$ initially points to a memory
location storing the identity function, and the function $f$ depends on $r$.
A reference pointing a memory location uses the arrow $|->$,
with the location and contents denoted by a box.
Code dependent on a reference uses the arrow $->$ to distinguish from pointing
to memory locations.
\[
\begin{tikzpicture}[
squarednode/.style={rectangle, draw=black, very thick, minimum size=5mm},
]
  \node[]                 (ref)                       {$r$};
  \node[]                 (fun)        [left=of ref]  {$\<lam> x . (\textbf{!}~ r) \ x$};
  \node[squarednode]      (cell)       [right=of ref] {$\<lam> x . x$};

  \draw[->]  (fun.east) -- (ref.west);
  \draw[|->] (ref.east) -- (cell.west);
\end{tikzpicture}
\]

After updating $r$ to $f$, the problematic cycle comes from the combination of the
dependency arrow for $f$ on $r$ and $r$ pointing to $f$.
\[
\begin{tikzpicture}[
squarednode/.style={rectangle, draw=black, very thick, minimum size=5mm},
]
  \node[]                 (ref)                       {$r$};
  \node[squarednode]      (cell)       [right=of ref] {$\lambda x . (\textbf{!}~ r) \ x$};

  \draw[|->] (ref.east) -- (cell.west);
  \draw[->]  (cell.north) to [out=150,in=30] (ref.north);
\end{tikzpicture}
\]

To prevent such cycles, we need to prevent $r$ from containing functions that
\emph{depend on} $r$.
The first diagram had no cycles, and updates to $r$ with other functions that are not
like $f$ has no such cycles: for example, updating $r$ to contain the
function $\lambda x . 0$.
\[
\begin{tikzpicture}[
squarednode/.style={rectangle, draw=black, very thick, minimum size=5mm},
]
  \node[]                 (ref)                       {$r$};
  \node[]                 (fun)        [left=of ref]  {$\lambda x . (\textbf{!}~ r) \ x$};
  \node[squarednode]      (cell)       [right=of ref] {$\lambda x . 0$};

  \draw[->]  (fun.east) -- (ref.west);
  \draw[|->] (ref.east) -- (cell.west);
\end{tikzpicture}
\]

To use a type universe hierarchy for preventing cycles,
we consider $r$ and functions with \emph{any dependencies} on $r$ to be in the
same universe.
The values that $r$ can store will be at a separate universe, so that
$r$ cannot be updated with any value from $r$'s universe,
preventing cyclic dependencies.
A type universe hierarchy similarly prevents circular reasoning in proofs encoded in
dependent type theory, where $\<Type>_i$ is considered to be of type $\<Type>_{i+1}$,
and not of type $\<Type>_i$.

To prevent other potential cycles, we also have to prevent $r$ from pointing
to any expressions in a \emph{higher} universe than $r$ since such expressions
could depend on $r$.
This additional restriction imposes an acyclic heap by stratifying the heap such that
references can only point ``down'' in the heap.

\begin{center}
  \includegraphics{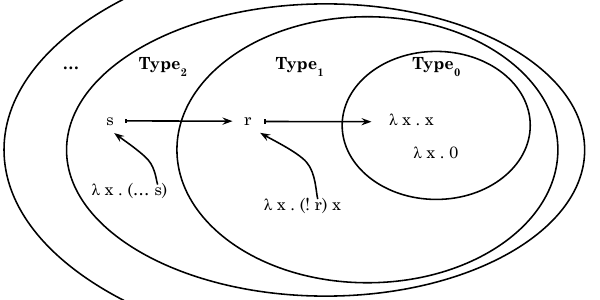}
\end{center}

This diagram may cause d\'ej\`a vu for those familiar with models of languages with
mutable references, where a similar diagram is also illustrated by
Ahmed~\cite{ahmedthesis}.
The stratification introduced in Ahmed's work is used in the \emph{model} to break the
type-world circularity, which we discuss in more detail in \Cref{ssec:termination}.
We reflect this stratification into the \emph{syntax} with a type universe hierarchy
for a stratified (and thus acyclic) heap.
Since functions are also allocated on the heap, our diagram here also includes
functions and their dependencies on the heap.

The key kinding rule in our type universe hierarchy for a stratified heap is
the rule for determining the universe level of reference types.
To require references to point ``downward'' in the heap, the universe level of a
reference type must be one level higher than the type it stores.
Here we use $::$ to indicate kinding a type.
\begin{mathpar}
  \inferrule
        {\tau :: \<Type>_i}
        {\refty{\tau} :: \<Type>_{i+1}}
\end{mathpar}
\noindent Given a type $\tau$ in universe $\<Type>_i$, then the type
$\refty{\tau}$ is in universe $\<Type>_{i+1}$.
For example, given $id$'s type is in universe $\<Type>_0$,
then $r$'s type is in universe $\<Type>_1$.

To complete the design of these stratified higher-order references,
we also consider how to determine the universe level of a function type,
since functions are also allocated on the heap.
The typical predicative function kinding rule in dependent type theories is
essentially as follows, where a function's universe level is the maximum of the
universe levels of its input and output types.
\begin{mathpar}
  \inferrule
        {\tau_1 :: \<Type>_i \\
        \tau_2 :: \<Type>_j \\
        k \geq i, j}
        {\tau_1 \to \tau_2  :: \<Type>_k}
\end{mathpar}

However, since functions are allocated as closures with code \emph{and} environment,
the universe information of the environment is lost with this kinding rule.
Without considering a function's environment, the functions $id$ and $f$ are
in the same universe.
But having $id$ and $f$ in the same universe violates the stratified heap
invariant, as $f$'s closure contains $r$.
The two functions $id$ and $f$ differ as \emph{closures}, since $id$ requires no
environment, but the function $f$'s environment depends on the reference $r$.
This means that the function $id$ and $f$, despite having the same input and
output types, will be in different universes based on their \emph{environment}.
The universe level of a function type is not only influenced by the
input and output types, but also the types captured in its environment!

To implement this, we include an annotation on the function arrow.
When viewing a function type in isolation, the only information available is the
universes of the input and output types, and not the (maximum) universe of the
environment.
We check that the universe annotation is consistent with the universe of the
input and output types.
We also separately check the annotation against the environment when the
type is used to check a term.
\begin{mathpar}
  \inferrule
        {\tau_1 :: \<Type>_i \\
        \tau_2 :: \<Type>_j \\
        k \geq i, j}
        {\funty{\tau_1}{\tau_2}{k}  :: \<Type>_k}
\end{mathpar}

The types $\funty{\<Nat>}{\<Nat>}{1}$ and $\funty{\<Nat>}{\<Nat>}{0}$ are both
well kinded, but describe different kinds of functions: the former are
permitted to capture references in universe $\<Type>_1$, while the latter cannot
capture references at all.
In our example, $f$'s type would be $\funty{\<Nat>}{\<Nat>}{1}$ since it captures
the universe $\<Type>_1$ reference $r$.
Then $r$ cannot be updated to $f$, since $f$ is in the same universe $\<Type>_1$,
preventing the cycle.

\FigSyntax
\FigTyping
\FigInfer
We now present the full language \lang, where PR stands for
\emph{predicative references}, inspired by predicative universe hierarchies.
The type system is essentially standard, with a type universe hierarchy to enforce
the heap stratification.
The type system is syntax directed, as is the kinding system, creating a simple,
declarative system for higher-order references with termination without complicated
resource tracking.

In \Cref{fig:syntax}, we present the syntax of \lang,
a simply typed $\lambda$-calculus with higher-order references.
The language includes base types, $\<Nat>$ and $\<Unit>$, denoted by the
metavariable $n$ and expression $\unite$ respectively.
The rest of the syntax is standard, with $\<nsnew>$, $\<nsderef>$, and $\coloneq$ for
initializing, dereferencing, and updating references.
The only exception is the annotation on the function type $\funty{\tau}{\tau}{k}$,
where $k$ indicates the universe of the function type and can be read as where
to allocate a function of this type on the heap.

In \Cref{fig:typing}, we present the typing rules of \lang.
The rules are standard, except for the function case.
Here we check (or could infer) the annotation on the function type against the
current context, $k \geq \<max>(\Gamma,\tau_1,\tau_2)$.
This side condition requires that the level $k$ of the function type is greater
than or equal to the levels of variables in $\Gamma$ and the levels of
types $\tau_1$ and $\tau_2$.
Note that the context $\Gamma$ could
contain more variables (and thus universe levels) than the function actually captures,
but through weakening the context one can easily determine the smallest level of a
function type by only the variables captured.

The function type annotation relies on kinding the input and output types and
the types in $\Gamma$ to determine the maximum universe level.
We present the kinding of types in \Cref{fig:kind}.
The simple base types $\<Nat>$ and $\<Unit>$ are of $\<Type>_0$ as expected.
The universe level of a function is determined by the annotation, given that this
annotation is greater than or equal to the input and output type universe levels.
And finally, a reference type is one level higher than the level of the
type it stores.

In \lang, our example of Landin's Knot is not well typed, which
we demonstrate through type and kind annotations rather than a large derivation tree.
\begin{displaymath}
  \begin{array}[t]{lll}
    id = (\<lam> x . x)  & : \funty{\<Nat>}{\<Nat>}{0} \\
    r = \<new> id & : \<Ref> (\funty{\<Nat>}{\<Nat>}{0}) :: \<Type>_1\\
    f = (\<lam> x . (\<deref> r) \ x) & : \funty{\<Nat>}{\<Nat>}{1} :: \<Type>_1 & \text{$\<Type>_1$ due to $r$}\\
    r := f & \text{type error, expected: $\funty{\<Nat>}{\<Nat>}{0}$} & \text{actual: $\funty{\<Nat>}{\<Nat>}{1}$} 
  \end{array}
\end{displaymath}

\subsection{Proof of Termination}
\label{ssec:termination}

The proof of termination for \lang uses the standard logical relations
technique, an introduction of which can be found in standard textbooks~\cite{atapl-lr}.
Our relation models types as sets of normalizing expressions.
We prove all well-typed expressions in \lang are in the set associated
with their type in \Cref{thm:fundamental}.
To be in the set associated with its type, an expression must step to a value, which
allows us to conclude that all well typed expressions step to a value, \ie terminate.

To model a language with mutable references, one has to model the heap, since
expressions access and modify the heap.
The model of a heap is often referred to as a \emph{world}, and is modelled as a
finite map from locations to sets of values modelling types.
Since each location is mapped to a set of all possible values, the world
represents any possible concrete execution heap.

However, we run into a circularity in reasoning when modelling heaps as worlds, known
as the \emph{type-world circularity}.

\begin{displaymath}
  \begin{array}{rcll}
    \textit{Type} & = & \textit{World} \rightarrow \textit{Set of Terms}\\
    \textit{World} & = & \textit{Loc} \rightharpoonup \textit{Type}
  \end{array}
\end{displaymath}

\noindent To model a type, we take in a world and produce a set of terms representing
that type.
The world maps locations to types and must also take in a world,
resulting in the following equation with an inconsistent cardinality, that is, the
set must contain itself: $\textit{World} = \textit{Loc} \rightharpoonup (\textit{World} \rightarrow \textit{Set of Terms})$.

\emph{Step-indexed logical relations} were introduced by Ahmed~\cite{ahmedthesis}
to model languages with mutable references, where the type-world circularity is
eliminated using a decreasing number $k$.
Using step-indexing, the equations become:

\begin{displaymath}
  \begin{array}{rcll}
    \textit{Type}_{k+1} & = & \textit{World}_{k} \rightarrow \textit{Set of Terms}\\
    \textit{World}_{k} & = & \textit{Loc} \rightharpoonup \textit{Type}_{k}
  \end{array}
\end{displaymath}

\noindent Unrolling $\textit{World}_k$, we have the following consistent equation
$\textit{World}_{k} = \textit{Loc} \rightharpoonup (\textit{World}_{k-1} \rightarrow \textit{Set of Terms}$.
In previous work by Ahmed, the decreasing metric $k$ is determined by
the number of steps of reduction that an expression takes.
Given an expression $e \in \textit{Type}_{k+1}(W_{k})$, the expression $e$ steps
``safely'', \ie without type errors, for $k+1$ steps.
Ahmed hints that there is a way to stratify types based on syntax, but chooses this
alternative semantic approximation in order to model arbitrarily quantified types.

We take the opposite approach to Ahmed and stratify the types based on syntax.
The decreasing metric in the type-world equations is instead based on the
universe level of a type.
The resulting equations are also slightly different, where the universe level matches
the world level, and a world level is increased by one based on the universe level of
the mapped types.

\begin{displaymath}
  \begin{array}{rcll}
    \textit{Type}_{i} & = & \textit{World}_{i} \rightarrow \textit{Set of Terms}\\
    \textit{World}_{i+1} & = & \textit{Loc} \rightharpoonup \textit{Type}_{i}
  \end{array}
\end{displaymath}

\noindent Unrolling $\textit{World}_{i+1}$, we have the following consistent equation
$\textit{World}_{i+1} = \textit{Loc} \rightharpoonup (\textit{World}_{i} \rightarrow \textit{Set of Terms})$.
These equations correspond closely to our kinding rule for references,
where reference type $\refty{\tau}$ is of $\<Type>_{i+1}$, given $\tau$ is $\<Type>_i$.
The type $\<Type>_{i+1}$ has access to $\textit{World}_{i+1}$, that is, locations mapping to
$\<Type>_i$, just as the reference at $\<Type>_{i+1}$ has access to $\tau$
at $\<Type>_i$.
One may also notice the absence of $\textit{World}_0$, where locations would be mapped to
$\<Type>_{-1}$, and this corresponds to the idea that values at $\<Type>_0$ do not
need access to the heap.
Another interpretation is that these values are valid in any world, because they
do not access or update the heap, and thus any world at any level
can stand in for $\textit{World}_0$.

\FigVR
Since the decreasing metric is included in the syntax, the resulting
logical relation reads as a simpler version of a step-indexed logical relation,
since each expression no longer needs to be paired with a step index $k$.
In \Cref{fig:vr}, we present the value relation for reference and function types.
The resulting sets are simply sets of values.
The value relation indexes the \emph{world} at the same level as the
universe, avoiding the type-world circularity and ensuring that values are
not accessing locations beyond their world level. 

The set of values for a reference type $\refty{\tau}$ is indexed by a number $i+1$,
since the universe level of $\refty{\tau}$ is always $i + 1$ for $\tau$'s level $i$.
The world associated with the type $\refty{\tau}$ is also at level $i + 1$.
Then, the set of values for type $\refty{\tau}$ is all the locations in the current
world that map to the set of values associated with $\tau$,\ie
$\mathcal{V}|[ \tau |]_{i}$.
The level of worlds associated with such values is necessarily lower than that
of the current world $W_{i+1}$, since the set is indexed by $i$.

\FigAuxDefs
The set of values for a function type $\funty{\tau}{\sigma}{k}$ relies on
additional definitions presented in \Cref{fig:ad}.
The first definition is \emph{world extension} $\sqsupseteq_k$, which
describes a \emph{future world} $W'_k$ with respect to the world $W_k$.
World extension is necessary because a function can be applied later in a
future \emph{heap}, and so the model must include function values valid in
future \emph{worlds}.
The relation guarantees that $W_k'$ has as many
locations as $W_k$ at the same types, but may have additional locations allocated.

The next definition \emph{lowers} a world $W_k$ to level $i$.
Lowering a world is necessary since the values in the relation
$\mathcal{V}|[ \tau |]_{i}$ rely on a world indexed at the same level $i$.
However, the available world $W'_k$ is at level $k$, which is potentially
incompatible with level $i$.
We use the lower operation to remove all locations that map to types with levels
\emph{higher} than level $i-1$.
The resulting world $\lw{W_k'}{i}$ is a $\textit{World}_i$ since
there are no locations mapping to types with level greater than $i-1$.
This lower operation can be considered a form of semantic garbage collection.

The final definition \emph{lifts} a world $W_k$ to a world that is
an intersection of worlds at all possible levels, denoted as
$\mathbf{W} = \forall i . \textit{World}_i$.
Such a world $\mathbf{W}$ is used in the expression relation, since an expression
may allocate and access levels higher than its universe level $i$.
The lift operation can be considered a ``cast'' from a world with types at
universe level $k-1$ allocated to a world that can allocate at higher levels.
The current allocated locations in the resulting world are the same as
$W_k$, and any other locations at higher levels than $k$ have yet to be
allocated, hence denoted as $\emptyset$.

The value relation is defined over a fully annotated function type, where the
universe levels of both the input type $\tau$ and output type $\sigma$ are known.
Kinding types is easy as shown in \Cref{fig:kind}, and the levels are
necessary for indexing the relations for $\tau$ and $\sigma$ correctly.
The set contains functions that, given any value $v$ in the value relation for the
input type $\tau$, the body of the function $e$ with the parameter $x$ substituted
with $v$ is in the \emph{expression} relation for $\sigma$.



\FigER

The expression relation defined in \Cref{fig:er} describes the set of terminating
expressions associated with a type $\tau$ at universe level $i$.
At a high level, the set consists of expressions that step to a value $v$ that is
in the value relation for $\tau$ at level $i$, that is $\mathcal{V}|[ \tau |]_{i}$.
The expression relation is defined over a world $\mathbf{W}$ at any level because
during evaluation, an expression may allocate and access levels higher than its
universe level $i$, as long as the final value does not depend on them.

There are three worlds $\mathbf{W}$, $\mathbf{W'}$, and $\mathbf{W''}$
in the expression relation.
The world $\mathbf{W}$ is considered the model of the minimum or initial
concrete heap needed for evaluation.
However, expressions can be evaluated in heaps larger than the initial heap, which
is why the expression relation includes the future world $\mathbf{W'}$.
Then, evaluation is under a concrete heap $h'$ realizing world $\mathbf{W'}$,
denoted by $\hwt{h'}{\mathbf{W'}}$.
The concrete heap $h'$ contains the same locations as $\mathbf{W'}$,
but maps each location to a \emph{single} value from the value relation mapped
by $\mathbf{W'}$.
There is a final concrete heap at the end of evaluation $h''$, and there must exist
a future world $\mathbf{W''} \sqsupseteq \mathbf{W'}$ related
to $h''$.
Finally, the value $v$ resulting from evaluation is in the value relation for
type $\tau$, with the final world $\mathbf{W''}$ lowered to $i$ since $v$ does not
rely on any part of the heap higher than $i$.
Lowering also maintains the stratification invariant since values such as
functions will be guaranteed not to access the heap at levels greater than $i$,
and thus can be ``deallocated''.

We prove \lang terminating by proving the fundamental lemma, which relies on a
substitution $\gamma$ respecting the context $\Gamma$, as defined in \Cref{fig:er}.
\begin{theorem}
  \label{thm:fundamental}
  If  $\tyjudg{\Gamma}{e}{\tau}$ and $\tau :: \<Type>_i$, then
  $\forall \gamma \in \mathcal{G}|[ \Gamma |](\mathbf{W}) . \gamma(e) \in \mathcal{E}|[ \tau |]_i(\mathbf{W})$.
\end{theorem}
This theorem states that if an expression $e$ is well typed at type $\tau$ with
universe level $\<Type>_i$, then $e$ is in the expression relation for $\tau$, \ie
$e$ steps to a value $v$ at type $\tau$. 
This theorem also applies to open terms by using a substitution $\gamma$.

The key takeaways from this proof is that a syntactic hierarchy of types is able to
resolve the type-world circularity by ensuring that there are no cycles in the heap,
and such a language is terminating.
We need a notion of garbage collection with the lower operation $\lw{-}{i}$,
which potentially allows for a language design where either the garbage collection
is explicit in the syntax, or can be inferred by the type system similar to
region type-and-effect systems.

\subsection{Interpretations and Our Framework}
We conclude our presentation of \lang with some intuitive
interpretations of the type system, although these connections are not yet
formalized.
We also present our general framework for instantiating the type system with
alternative type universe hierarchies, and present additional instantiations in
subsequent sections.

One interpretation of \lang is as a region type-and-effect system, where each
universe level is considered as one very large region.
All pure values are in region (level) 0, references to level 0 and functions that
close over references are in region (level) 1, and so on.
Instead of separate typing and kinding judgments, one could perhaps combine them
into one type-and-effect judgment as $\Gamma \vdash e : (\tau , \textbf{region } i)$,
where the effect $\textbf{region } i$ describes allocation in region $i$ and is
determined by kinding $\tau$.
Additionally, if a program is made up of values that do not exceed
region $n$ (level $n$), then any regions above $n$ can be safely deallocated.

The universe levels can also be interpreted as an allocation effect.
Values of types in universe $\<Type>_0$ are allocated at region 0 of the heap.
Values of types in universe $\<Type>_1$ are allocated in region 1 of the heap, where
values have access to region 0, but not any regions above, and so on
for each additional region.
Simply based on the type, one knows where a value of a certain type
will be allocated on the heap.
The structure of the heap is maintained for all programs because values are
allocated in their appropriate regions, as enforced by the type hierarchy.

Where a function is allocated is dependent on what parts of the heap the function
relies on.
Interestingly, a function may allocate and update \emph{new} references and still
be considered at $\<Type>_0$, \ie allocated in region 0, for example:
\begin{displaymath}
  \begin{array}[t]{l}
    ex :  \funty{\<Nat>}{\<Nat>}{0} \\
    ex = (\<lam> x . r = \<new> 3 ; !r \ + \ x)
  \end{array}
\end{displaymath}
One can consider the reference $r$ as completely ``private'' to the function $ex$,
and $r$ does not affect the existing structure of the heap nor does $r$ ever leave
the function's scope.
Under this effect system, functions that do not leak effectful values can be considered
``pure'' in the sense they do not depend on any part of the heap and thus can be used in
any arbitrary heap.
Furthermore, the allocations performed inside $ex$ are essentially invisible after
the computation ends, and can freely be collected or optimized away.

There may also be an interpretation of \lang as a coeffect system~\cite{petricek2014},
since the function typing rule is dependent on the context.
In coeffect systems, a function type is annotated with an effect derived from
the current context, and variables in the context are annotated with their effects.
A possible coeffect judgment for function typing is as follows,
assuming $\tau$ has level $i$.
\begin{mathpar}
  \inferrule{\Gamma , x :^i \tau \vdash e : \sigma \\ k = \textbf{max-effect}(\Gamma)}
            {\Gamma \vdash \lam{x}{\tau}{e} : \funty{\tau}{\sigma}{k}}
\end{mathpar}

For our more general framework, we abstract the design of
\lang's type system into a few ``knobs'' that can be adjusted to create different
designs.
The abstraction leads to two parameters that can be adjusted between designs.
\begin{enumerate}
\item The highlighted side condition for typing a function, in \lang this equation is $k \geq \<max>(\Gamma,\tau_1,\tau_2)$.
\item The relationship between the universe level of a reference type and the type it stores. In \lang, a reference universe level was always one level higher.
\end{enumerate}
What these equations create is essentially the desired algebra
of the universe levels, \ie the values of the levels and operations over them.
In \lang, the algebra consists of natural numbers, a $\geq$ operator,
and a successor operator for the level of reference types.

We now explore how to adjust these parameters for alternative designs.

\section{A Universe Hierarchy with Impredicativity}

One type universe hierarchy already in use in dependent type theories
allows one universe level to be \emph{impredicative}, and the rest of the levels are
predicative~\cite{luo1990}.
The impredicative level allows propositions to quantify over their own level, \ie
propositions can be sound with cyclic definitions, with some limitations.
We conjecture similar reasoning can be used to distinguish between ``sound'' and
``unsound'' cycles in the memory heap.
Many data structures are encoded using cycles, so we believe there is a
correspondence between ``sound'' cycles for data and ``unsound'' cycles
resulting in nontermination.

We conjecture that \emph{full-ground} references~\cite{murawski2012}, \ie references
to base data and other full-ground references, can be encoded using an
impredicative universe level.
The idea is that references containing base types will remain at the same level as
the base universe $\<Type>_0$.
Functions are not included as base types for full-ground references, and will be
at universe $\<Type>_1$ and above.
When a reference stores a function, the reference is no longer full-ground, and
is subject to the stratified (predicative) part of the universe hierarchy.

Adding an impredicative universe level is easily done by tweaking our two parameters.
For the function case, the equation is changed from $\geq$ to $>$, resulting in a
system where pure functions and functions that close over full-ground references
are at $\<Type>_1$.
Functions that close over other higher-order references continue up the
universe hierarchy similarly to \lang.
\begin{mathpar}
  \inferrule
        {\tyjudg{\Gamma,x:\tau_1}{e}{\tau_2} \\
        \colorbox{gray}{$k > \<max>(\Gamma,\tau_1,\tau_2)$}}
        {\tyjudg{\Gamma}{\lam{x}{\tau_1}{e}}{\funty{\tau_1}{\tau_2}{k}}}
\end{mathpar}
Without this change to $>$, Landin's Knot would be well typed, since both the function
$id$ and the function that closes over a reference $f$ would be in the same universe.

For reference types, the universe of a reference type now depends on the level
of what it stores.
When storing a type at base universe $\<Type>_0$, the reference universe level remains
the same.
For any types above level 0, the reference universe level is one higher.
\begin{mathpar}
  \inferrule
        {\tau :: \<Type>_0}
        {\refty{\tau} :: \<Type>_{0}}
        \qquad
  \inferrule
        {\tau :: \<Type>_i \\ i \neq 0}
        {\refty{\tau} :: \<Type>_{i+1}}
\end{mathpar}
\noindent The resulting type system has full-ground references with sound cycles at
$\<Type>_0$, but unsound cycles that result in non-termination are still prevented
by the stratification imposed by predicativity.
We conjecture that this system is still terminating and has the ability to create
cyclic data.

The resulting system will likely have a stratified structured heap as seen in \lang,
with cycles occurring in level 0 of the heap.

\begin{center}
  \includegraphics{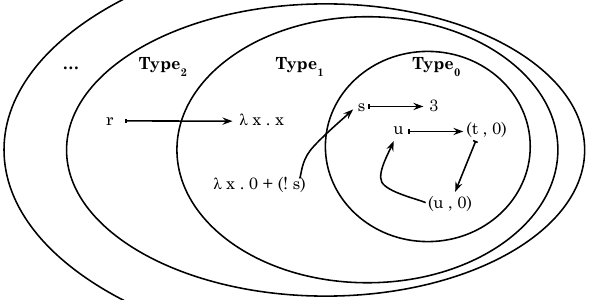}
\end{center}

\noindent Here we have a $\<Type>_0$ level of the heap where base data can be
allocated along with references to base data, and thus can encode a cyclic list.

\section{Type Universe Level Polymorphism}

Abstracting explicit levels has been the subject of many research papers on
the design of type universe hierarchies.
\emph{Universe polymorphism} allows expressions to be polymorphic with respect to
universe levels, to eliminate the redundancy of defining the same expressions at
different universe levels.
Current implementations of dependent type theories use three main techniques
for universe polymorphism: explicit quantification for universe levels
as done in Agda~\cite{agda:universes},
user constraints as done in Coq~\cite{coq:explicituniverses}, and
constraint solving based on \emph{typical ambiguity} originally developed by
Huet~\cite{huet1988} and Harper and Pollack~\cite{harper1991} and also
implemented in Coq.
Recent work by Hou et al.~\cite{hou2023} influenced by
McBride's \emph{crude-but-effective stratification}~\cite{mcbride:cbe-strat} has
found that an alternative system with an explicit displacement operator is
a simple and effective mechanism for universe polymorphism.

We conjecture that universe polymorphism allows us to ``name'' regions of the heap,
resulting in a type system similar to region type-and-effect systems.
To implement McBride's \emph{crude-but-effective stratification} version of universe
polymorphism, our language needs an explicit displacement operator on both expressions
and types. 
Then, a value typically in one ``region'' of the heap can be lifted to another
region using the displacement operator.

Most base values exist in region 0 (\ie in universe $\<Type>_0$), and having
such a large region that can never be deallocated is poor for fine-grained memory
management.
We then use the displacement operator to distinguish values into different
regions.
For example, given a term $e$ of type $\tau$, suppose we would like to allocate
$e$ not in region 0 but in some region we call $\beta$.
We use the explicit displacement operator $\Uparrow_\beta$ to do this,
an operator also defined over types that updates the kind level accordingly.

\begin{mathpar}
  \inferrule
      {\Gamma \vdash e : \tau}
      {\Gamma \vdash \Uparrow_\beta e : \Uparrow_\beta \tau}
      \qquad
  \inferrule
        {\tau :: \<Type>_\alpha}
        {\Uparrow_\beta \tau :: \<Type>_{\alpha+\beta}}
\end{mathpar}

Without adjusting the equations \lang, we can now allocate
values in different levels of the stratified heap.

\begin{center}
  \includegraphics{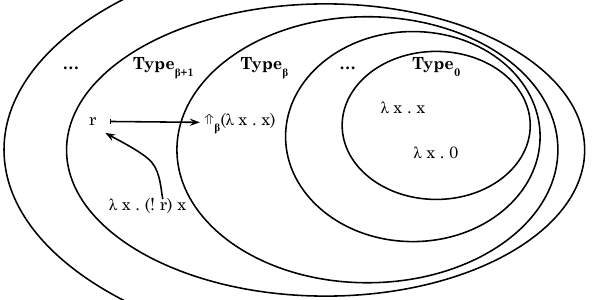}
\end{center}

\noindent When no expression relies on a particular region (level) $\beta$,
any values in and references to level $\beta$ can be safely deallocated.
Higher levels are considered separate regions, and allow for fine-grained
region-like memory management.
The advantages of this system with a lifting operator is that the system is still
completely syntax directed, whereas some region type-and-effect systems need to
infer the regions based on reads and writes.

However, because we haven't changed the equations from \lang,
the resulting regions still have a stratified structure.
We conjecture that a more advanced system is possible, where the relationship
between a reference universe level and the universe level it stores can be
specified through some other relation besides successor.
\begin{mathpar}
  \inferrule
        {\tau :: \<Type>_\alpha \\ \mathcal{R}(\alpha,\beta)}
        {\refty{\tau} :: \<Type>_{\beta}}
\end{mathpar}

To obtain the stratified variant, $\mathcal{R}$ specifies that $\beta = \alpha+1$.
However, to obtain different heap shapes, for example a shape where values allocated
in region $\alpha$ are completely separate from references to $\alpha$, we could
specify $\alpha \ \# \ \beta$ to mean that $\alpha$ and $\beta$ are disjoint.
Alternatively, certain values and references to these values could be allocated
in the same region by specifying $\alpha = \beta$.
With this abstraction, different heap shapes could be formed simultaneously.
The pictured heap below has distinct regions $\alpha$, $\beta$, and $\gamma$, where
references in $\alpha$ are stratified, references in $\beta$ are in $\beta$, and
$\gamma$ exists to contain references separate from values in $\alpha + 1$.

\begin{center}
  \includegraphics{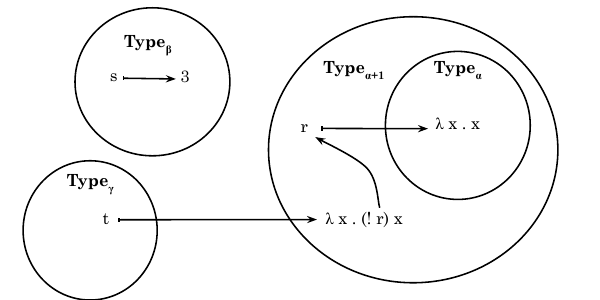}
\end{center}

This kinding rule is no longer syntax directed, and
may result in a type universe polymorphism system more similar to Coq, where
the universes are inferred using constraint solving.
Alternatively, a user could declare universes (regions), and constraints between them.
For example, the following constraints could be specified to create the heap
visualized above, with the assumption that reference types bump the universe level by 1
when no constraints are given.
\begin{displaymath}
  \begin{array}[t]{l}
    \<Type>_{\alpha} \ \# \ \<Type>_{\beta} \ \# \ \<Type>_{\gamma} \\
    \<Type>_{\beta + 1} = \<Type>_{\beta} \\
    \<Type>_{\alpha + 2} = \<Type>_{\gamma} 
  \end{array}
\end{displaymath}
\noindent These constraints specify that three regions, $\alpha$, $\beta$, and $\gamma$
are all disjoint.
However, references to values in region $\beta$ are still in $\beta$,
hence the constraint $\<Type>_{\beta + 1} = \<Type>_{\beta}$.
Additionally, references to level $\alpha + 1$, \ie $\<Type>_{\alpha+2}$, will be
in $\<Type>_{\gamma}$, since the reference $t$ is in $\<Type>_{\gamma}$.

\section{Conclusion}
What we've studied in this paper is the connection between type universe hierarchies
and memory allocation.
By studying an example of non-termination using higher-order references, we found
a simple mechanism to prevent such cycles by equipping a type system with a
type universe hierarchy.
The universe hierarchy imposes a structure on the heap for all programs, and
we found that the kind of a type describes where values can be allocated on the heap.
We are able to distill the essence of this type system to two parameters,
one describing the relationship of a function's type level to the levels it captures,
and another describing the relationship between a reference type level and what
level it stores.
These parameters give us a theoretical framework to create type systems that
enforce different heap designs by changing the algebra of the universe hierarchy.
We believe the simplicity and power of these systems with a universe hierarchy
provide a fresh foundation for new language designs for low-level memory reasoning.

\section{Acknowledgments}

We are extemely grateful for Ohad Kammar's guidance, enthusiasm, and patience
while discussing this project, and the countless hours you spent trying
to teach us categorical semantics for local state, which helped us immensely
with our own semantics.
We are also grateful to Max S. New, who provided early intuitions into this work and
the model more generally, as well as a proof of termination for higher-order
references with \emph{closed} functions.
We are also thankful to Ugo Dal Lago for pointing us to the stratified regions work
~\cite{boudol2010,amadio2009}.
We also would like to thank all who engaged with our work in presentations at the
workshops HOPE and PNW PLSE, and other informal venues.
Your enthusiasm, suggestions, and critiques are all a part of our work more than
you know, and we appreciate your contribution to the scientific process.
This material is based upon work supported by the Defense Advanced Research Projects
Agency (DARPA) and Naval Information Warfare Center Pacific (NIWC Pacific) under
Contract No. NN66001-22-C-4027.
Any opinions, findings and conclusions or recommendations expressed in this material
are those of the author(s) and do not necessarily reflect the views of DARPA or NIWC
Pacific.

\bibliographystyle{splncs04}
\bibliography{all}

\begin{thebibliography}{10}
\providecommand{\url}[1]{\texttt{#1}}
\providecommand{\urlprefix}{URL }
\providecommand{\doi}[1]{https://doi.org/#1}

\bibitem{ahmed2007}
Ahmed, A., Fluet, M., Morrisett, G.: L3: A linear language with locations
  (2007). \doi{10.1007/11417170_22}

\bibitem{ahmedthesis}
Ahmed, A.J.: Semantics of Types for Mutable State. Ph.D. thesis (2004),
  \url{http://www.ccs.neu.edu/home/amal/ahmedthesis.pdf}

\bibitem{amadio2009}
Amadio, R.M.: On Stratified Regions. Asian Symposium on Programming Languages
  and Systems ({APLAS}) (2009). \doi{10.1007/978-3-642-10672-9_16}

\bibitem{boudol2010}
Boudol, G.: Typing termination in a higher-order concurrent imperative
  language. Information and Computation  (2010). \doi{10.1016/j.ic.2009.06.007}

\bibitem{atapl-lr}
Crary, K.: Logical relations and a case study in equivalence checking. In:
  Pierce, B.C. (ed.) Advanced Topics in Types and Programming Languages,
  chap.~6. MIT Press (2005)

\bibitem{harper1991}
Harper, R., Pollack, R.: Type checking with universes. Theoretical Computer
  Science  (1991). \doi{10.1016/0304-3975(90)90108-t}

\bibitem{hou2023}
Hou~(Favonia), K.B., Angiuli, C., Mullanix, R.: An order-theoretic analysis of
  universe polymorphism. In: Symposium on Principles of Programming Languages
  ({POPL}) (2023). \doi{10.1145/3571250}

\bibitem{huet1988}
Huet, G.: Extending the {Calculus of Constructions} with {Type:Type} (1988),
  \url{http://pauillac.inria.fr/~huet/PUBLIC/typtyp.pdf}

\bibitem{jutting1993}
Jutting, L.: Typing in pure type systems. Information and Computation  (1993).
  \doi{10.1006/inco.1993.1038}

\bibitem{landin1964}
Landin, P.J.: The mechanical evaluation of expressions. The Computer Journal
  (1964). \doi{10.1093/comjnl/6.4.308}

\bibitem{luo1990}
Luo, Z.: An Extended Calculus of Constructions. Ph.D. thesis, University of
  Edinburgh (1990),
  \url{http://www.lfcs.inf.ed.ac.uk/reports/90/ECS-LFCS-90-118/}

\bibitem{matsakis2014}
Matsakis, N.D., Klock, F.S.: The {Rust} language. In: High Integrity Language
  Technology (2014). \doi{10.1145/2663171.2663188}

\bibitem{mcbeth1963}
McBeth, J.H.: Letters to the editor: on the reference counter method.
  Communications of the ACM  (1963). \doi{10.1145/367593.367649}

\bibitem{mcbride:cbe-strat}
McBride, C.: Crude but effective stratification (2011),
  \url{https://mazzo.li/epilogue/index.html%3Fp=857&cpage=1.html}

\bibitem{murawski2012}
Murawski, A.S., Tzevelekos, N.: Algorithmic Games for Full Ground References.
  International Colloquium on Automata, Languages, and Programming ({ICALP})
  (2012). \doi{10.1007/978-3-642-31585-5_30}

\bibitem{petricek2014}
Petricek, T., Orchard, D., Mycroft, A.: Coeffects: a calculus of
  context-dependent computation. In: International Conference on Functional
  Programming ({ICFP}) (2014). \doi{10.1145/2628136.2628160}

\bibitem{coq:explicituniverses}
Sozeau, M.: Polymorphic universes ({Coq} reference manual),
  \url{https://coq.inria.fr/doc/V8.18.0/refman/addendum/universe-polymorphism.html#explicit-universes}

\bibitem{agda:universes}
{The Agda Development Team}: Universe levels ({Agda} reference manual),
  \url{https://agda.readthedocs.io/en/v2.6.4.3-r1/language/universe-levels.html}

\bibitem{tofte1997}
Tofte, M., Talpin, J.P.: Region-based memory management. Information and
  Computation  (1997). \doi{10.1006/inco.1996.2613}

\end{thebibliography}

\end{document}